\documentclass[12pt]{iopart}
\input{amssym}
\usepackage{epsfig}
\global\arraycolsep=2pt

\newcommand{\be}{\begin{equation}}
\newcommand{\bnabla}{\mbox{\boldmath{$\nabla$}}}
\newcommand{\mathfrak}{\frak}
\newcommand{\ee}{\end{equation}}
\newcommand{\bea}{\begin{eqnarray}}
\newcommand{\eea}{\end{eqnarray}}
\newcommand{\ben}{\begin{equation*}}
\newcommand{\een}{\end{equation*}}
\newcommand{\bean}{\begin{eqnarray*}}
\newcommand{\eean}{\end{eqnarray*}}

\begin{document}

\title{Surface divergences and boundary energies in the Casimir effect}
\author{KA Milton\footnote{on sabbatical
leave at the Department of Physics, Washington University, St. Louis,
MO 63130 USA},
I Cavero-Pel\'aez and J Wagner}
\address{Oklahoma Center for High Energy Physics and
Homer L. Dodge Department of Physics and Astronomy,
University of Oklahoma, Norman, OK 73019 USA}
\ead{milton@nhn.ou.edu}
\begin{abstract}
Although Casimir, or quantum vacuum, forces between
distinct bodies, or self-stresses of individual bodies,
have been calculated by a variety of different methods since
1948, they have always been plagued by divergences.
Some of these divergences are associated with the volume, and
so may be more or less unambiguously removed,
while other divergences are associated with the surface.
The interpretation of these has been quite controversial.
Particularly mysterious is the contradiction between finite
total self-energies  and surface divergences
in the local energy density.
In this paper we clarify the role of surface divergences.
\end{abstract}

\section{Introduction}

The subject of local energy density associated with the confinement of
quantum fields by surfaces has a rather long history.  For example,
Brown and Maclay \cite{Brown:1969na} computed the vacuum expectation
value of the electromagnetic
energy-momentum tensor between two parallel perfectly conducting plates, which 
is twice that of a conformally coupled 
massless scalar field satisfying Dirichlet or
Neumann boundary conditions on the plates, namely for plates separated
by a distance $a$ in the $x$ direction,
\be 
\langle T^{\mu\nu}\rangle=\frac{\pi^2}{1440 a^4}\mbox{diag}\,(-1,-3,1,1),
\ee
which corresponds precisely to the attractive energy or pressure found
by Casimir \cite{Casimir:1948dh}  in the same situation.
If a nonconformal
scalar stress tensor is used, a position-dependent term in the stress tensor
appears, which does not contribute to either the total energy or the pressure
on the plates \cite{Milton:2002vm,Actor:1996zj}.

Local surface divergences were first discussed for arbitrary smooth boundaries
by Deutsch and Candelas \cite{Deutsch:1978sc}.  They found cubic divergences
in the energy density as one approaches the surface; for example, outside
a Dirichlet sphere (that is, for a conformally-coupled
scalar field satisfying Dirichlet boundary
conditions on the surface) the energy density diverges as
\be
r\to a+:\quad u\sim \frac1{360\pi^2}\frac1{a(r-a)^3},\label{divdc}
\ee
where $a$ is the radius of the sphere.

\begin{figure}
\centerline{\psfig{figure=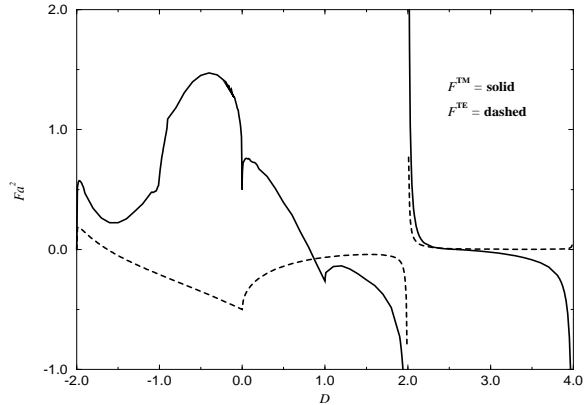,height=9cm,width=6cm,angle=270}}
\caption{A plot of the TM Casimir stress  $F^{\rm TM}$
for $-2<D<4$ on a spherical shell, compared with $F^{\rm TE}$, taken
from Bender and Milton \cite{Bender:1994zr,Milton:1996ri}.
For $D<2$ ($D<0$) the stress  $F^{\rm TM}$ ($F^{\rm TE}$)
is complex and we have plotted ${\rm Re}\,F$.}
\label{fig1}
\end{figure}

This raises the question: How can it be that the total Casimir energy of
a Dirichlet sphere (or a perfectly conducting sphere in electrodynamics)
is finite?  The electromagnetic case is the well-known one first calculated
by Boyer \cite{Boyer:1968uf},
$\mathcal{E}^{EM}=0.04618/a$,
while the scalar case was first
worked out by Bender and Milton \cite{Bender:1994zr},
$\mathcal{E}^S=0.002817/a$.
In general the Casimir energy of a region bounded by a perfect
hyperspherical surface depends in a complicated way upon the number
of spatial dimension $D$, as shown in Figure \ref{fig1}.

Thus there has been a suspicion since the time of Deutsch and Candelas 
that there was something incomplete in  the calculations of Casimir self
energies of ideal closed boundaries.
(We note that there is now a proof
that any such smooth perfectly conducting boundary possesses a finite 
electromagnetic Casimir energy \cite{graf}.
Whether such an idealized limit is physical is, of course, another question.) 
This suspicion has been
recently intensified by a series of talks and papers by Graham \etal
\cite{Graham:2003ib}.  
The essential outcome of their analysis is that
for a $\delta$-function sphere, described by the following Lagrangian
for a massless scalar field,
\be
\mathcal{L}=-\frac12\partial_\mu\phi\partial^\mu\phi-\frac12
\frac{\lambda}{a^2}\delta(r-a)\phi^2,
\ee
 a divergence occurs in third order in $\lambda$.
(They claimed a divergence in second order, but that was spurious,
arising from the omission of a surface term in the integration by parts
\cite{Milton:2002vm,Milton:2004ya,Milton:2004vy}--See \eref{totalenergy1}
below.)  This
divergence in $\Or(\lambda^3)$
in fact was discovered much earlier by Bordag, Kirsten, and
Vassilevich \cite{Bordag:1998vs}, and possible ways of dealing with it
have been suggested \cite{Bordag:2004rx,Scandurra:1998xa}.
Objections complementary to those of Graham \etal  have also
been voiced by Barton \cite{barton,barton2}, 
all of which raise doubts as to the
physical relevance of results such as those of Boyer.

\section{Green's function for $\lambda$ sphere}
We consider the potential
\be
\mathcal{L}_{\rm int}=-\frac\lambda{2a^2}\phi^2\sigma(r),
\quad
{\rm where}
\quad
\sigma(r)=\cases{
0,&$r<a_-$,\\
h,&$a_-<r<a_+$,\\
0,&$a_+<r$.}
\ee
Here $a_\pm=a\pm\delta/2$, and we set $h\delta=1$.  
We have
chosen the dimensions of $\lambda$ so that the total energy of
interaction does not explicitly refer to the radius $a$. In the limit
as $\delta\to0$ (or $h\to\infty$) we recover the $\delta$-function sphere.

The Green's function equation, with
$\kappa^2=-\omega^2$,
\be
\left(-\nabla^2+\kappa^2+\frac\lambda{a^2}
\sigma\right)\mathcal{G}(\mathbf{r,r'})=\delta(\mathbf{r-r'}),
\ee
may be straightforwardly solved.
We introduce the reduced Green's function
\be
\mathcal{G}(\mathbf{r,r'})=
\sum_{lm}g_l(r,r')Y_{lm}(\theta,\phi)Y_{lm}^*(\theta',\phi'),
\label{redgreen}
\ee
which in turn may be expressed in terms of
the modified Riccati-Bessel functions, 
\be
s_l(x)=\sqrt{\frac{\pi x}2}I_{l+1/2}(x),\quad
e_l(x)=\sqrt{\frac{2x}\pi}K_{l+1/2}(x).
\ee
The reduced Green's function is, outside of the shell:
\numparts
\bea
r,r'<a_-:\quad g_l&=&\frac1{\kappa rr'}\left[s_l(\kappa r_<)e_l(\kappa r_>)
-\frac{\tilde\Xi}{\Xi}s_l(\kappa r)s_l(\kappa r')\right],\label{gouta}\\
r,r'>a_+:\quad g_l&=&\frac1{\kappa rr'}\left[s_l(\kappa r_<)e_l(\kappa r_>)
-\frac{\hat\Xi}{\Xi}e_l(\kappa r)e_l(\kappa r')\right].\label{goutb}
\eea
\endnumparts
Here the denominator is
\bea
\fl\Xi=[
\kappa {s_l'(\kappa a_-)}
e_l(\kappa' a_-)-{\kappa' s_l(\kappa a_-)}
e_l'(\kappa'a_-)][\kappa' e_l(\kappa a_+)s_l'(\kappa'a_+)
-\kappa e_l'(\kappa a_+)s_l(\kappa'a_+)]\nonumber\\
\fl\quad\mbox{}-[{\kappa s_l'(\kappa a_-)}
s_l(\kappa' a_-)-\kappa'{s_l(\kappa a_-)}
s_l'(\kappa'a_-)]
[\kappa' e_l(\kappa a_+)e_l'(\kappa'a_+)-\kappa
e_l'(\kappa a_+)e_l(\kappa' a_+)],
\eea
while the numerator
$\tilde\Xi$ is obtained from $\Xi$ by replacing $s_l(\kappa a_-)\to
e_l(\kappa a_-)$, and $\hat\Xi$ is obtained from $\Xi$ by replacing
$e_l(\kappa a_+)\to s_l(\kappa a_+)$. 
Here $\kappa'=\sqrt{\kappa^2+\lambda h}$.
Green's function within  the shell, $a_-<r<a_+$, is given by
\bea\label{gin}
\fl g_l=\frac1{\kappa' rr'}\bigg\{s_l(\kappa'r_<)e_l(\kappa'r_>)
-\frac1\Xi\bigg[
[s_l(\kappa'r)e_l(\kappa'r')+s_l(\kappa'r')e_l(\kappa'r)]\nonumber\\
\fl\quad\times[\kappa e_l'(\kappa a_+)e_l(\kappa'a_+)-\kappa'e_l(\kappa a_+)
e_l'(\kappa'a_+)]
[\kappa s_l'(\kappa a_-)s_l(\kappa'a_-)
-\kappa's_l(\kappa a_-)s_l'(\kappa'a_-)]\nonumber\\
\mbox{}-s_l(\kappa'r')s_l(\kappa'r)[\kappa e_l'(\kappa a_+)e_l(\kappa'a_+)
-\kappa'e_l(\kappa a_+)e'_l(\kappa' a_+)]\nonumber\\
\quad\times[\kappa s_l'(\kappa a_-)e_l(\kappa' a_-)-\kappa' s_l(\kappa a_-)
e_l'(\kappa' a_-)]\nonumber\\
\mbox{}-e_l(\kappa'r')e_l(\kappa'r)[\kappa e_l'(\kappa a_+)s_l(\kappa'a_+)
-\kappa'e_l(\kappa a_+)s'_l(\kappa' a_+)]\nonumber\\
\quad\times[\kappa s_l'(\kappa a_-)s_l(\kappa' a_-)-\kappa' s_l(\kappa a_-)
s_l'(\kappa' a_-)]\bigg]\bigg\}.
\eea

\section{Energy density}
We can calculate the local energy density from the stress tensor:
\be
T^{\mu\nu}=\partial^\mu\phi\partial^\nu\phi-g^{\mu\nu}\mathcal{L}
-\xi(\partial^\mu\partial^\nu-g^{\mu\nu}\partial^2)\phi^2,
\ee
from which the energy density follows:
\be
T^{00}=\frac12\left[\partial^0\phi\partial^0\phi+\bnabla\phi\cdot\bnabla\phi
+\frac{\lambda}{a^2}\sigma\phi^2\right]-\xi\nabla^2\phi^2,
\ee
where the conformal value is given by $\xi=1/6$. 
To obtain the one-loop vacuum expectation values, we use the
connection to the Green's function
\be
 \langle\phi(x)\phi(x')\rangle=\frac1\rmi G(x,x').
\ee
The energy density thus is, within or outside the shell,
\bea
\fl \langle T^{00}\rangle&=&\frac1{2\rmi}\left(\partial^0\partial^{\prime0}+
\bnabla
\cdot\bnabla'+\left\{\begin{array}{c}\lambda h/a^2\\0\end{array}\right\}\right)
G(x,x')\bigg|_{x'=x}-\frac\xi{\rmi}\nabla^2G(x,x).
\eea
When we insert the partial wave decomposition of the Green's function
\eref{redgreen}, 
the expression for the energy density is immediately reduced
to (inside or outside the shell, but not within it)
\bea
\fl\langle T^{00}\rangle =\int_0^\infty \frac{\rmd\kappa}{2\pi}
\sum_{l=0}^\infty
\frac{2l+1}{4\pi}\!\left\{\!
\left[-\kappa^2+\partial_r\partial_{r'}+\frac{l(l+1)}{r^2}
\right]g_{l}(r,r')\bigg|_{r'=r}-2\xi\frac1{r^2}\frac{\partial}{\partial r}
r^2\frac{\partial}{\partial r}g_l(r,r)\right\}.\nonumber\\
\eea

We insert the Green's function in the exterior region, but delete the
free part, the first term in (\ref{gouta}), \eref{goutb}, (\ref{gin}),
which corresponds to the {\em bulk energy\/} which would be
present if either medium filled all of space, leaving us with for $r>a_+$
(for $r<a_-$, $e_l\to s_l$ and $\hat\Xi\to\tilde \Xi$)
\bea
\fl  u(r)=-(1-4\xi)\int_0^\infty \frac{\rmd\kappa}{2\pi}\sum_{l=0}^\infty 
\frac{2l+1}{4\pi}\frac{\hat\Xi}{\Xi}\bigg\{\frac{e_l^2(\kappa r)}{\kappa r^2}
\bigg[-\kappa^2\frac{1+4\xi}{1-4\xi}
+\frac{l(l+1)}{r^2}+\frac1{r^2}
\bigg]\nonumber\\
\mbox{}-\frac2{r^3}e_l(\kappa r)e_l'(\kappa r)+\frac\kappa
{r^2}e_l^{\prime2}(\kappa r)\bigg\}.
\eea

\section{Surface divergences}
We want to examine the singularity structure as $r\to a_++$.   For this purpose
we use the leading uniform asymptotic expansion, $l\to\infty$,
\bea
\fl  e_l(x)&\sim& \sqrt{zt}\,\rme^{-\nu \eta},\quad
s_l(x)\sim\frac12\sqrt{zt}\,\rme^{\nu \eta},\quad
e_l'(x)\sim-\frac1{\sqrt{zt}}\,\rme^{-\nu\eta},\quad s'_l(x)\sim
\frac12\frac1{\sqrt{zt}}\,\rme^{\nu \eta},
\eea
where $\nu=l+1/2$,
$x=\nu z$, $t=(1+z^2)^{-1/2}$, $\rmd\eta/\rmd z=1/{zt}$.

Let us consider the thin shell limit, $\delta\to 0$, $h\delta=1$, where
it is easy to check that
\be
\frac{\hat\Xi}{\Xi}\to\frac{\frac\lambda{\kappa a^2} s^2_l(\kappa a)}
{1+\frac\lambda{\kappa a^2}e_l(\kappa a)s_l(\kappa a)},
\ee
which is exactly the coefficient occurring in the $\delta$-function 
potential.  There are two simple limits of this, strong and weak coupling:
($\kappa a\sim 1$)
\bea
\frac{\lambda}a \to\infty:\quad\frac{\hat\Xi}{\Xi}\to
\frac{s_l(\kappa a)}{e_l(\kappa a)},\qquad
\frac{\lambda}a\to0:\quad  \frac{\hat\Xi}{\Xi}\to
\frac\lambda{\kappa a^2}s_l^2(\kappa a).
\eea
In either case, we carry out the asymptotic sum over angular momentum using
the uniform asympotic expansion  and
\be
\fl \sum_{l=0}^\infty \rme^{-\nu\chi}=\frac1{2\sinh\frac\chi2},\qquad
\chi=2\left[\eta(z)-\eta\left(z\frac{a}r\right)\right]\approx
2z\eta'(z)\frac{r-a}r=\frac2t\frac{r-a}r.
\ee
The remaining integrals over $z$ are elementary, and in this 
way we find that the leading divergences are as $r\to a+$,
\numparts
\bea
 \frac\lambda{a}\to\infty:\quad u&\sim&
-\frac1{16\pi^2}\frac{1-6\xi}{(r-a)^4},\label{dsphere1}\\
\frac\lambda{a}\to0:\quad  u^{(n)}&\sim&
\left(-\frac\lambda{a}\right)^n\frac{\Gamma(4-n)}{96\pi^2a^4}(1-6\xi)
\left(\frac{a}{r-a}\right)^{4-n},\quad n<4,
\eea
\endnumparts
the latter being the leading divergence in order $n$, 
which clearly seems to demonstrate the virtue of the conformal value of 
$\xi=1/6$.
(The  value for the Dirichlet sphere first appeared
in Deutsch and Candelas \cite{Deutsch:1978sc}.)
Thus, for $\xi=1/6$ we must keep subleading terms. 
This includes keeping the subdominant term in $\chi$,
and the distinction between $t(z)$ and $\tilde t=t(\tilde z=za/r)$,
\be
\chi\approx\frac2t\frac{r-a}r-t\left(\frac{r-a}r\right)^2,\qquad
\tilde z\tilde t\approx zt-t^3z\frac{r-a}r,
\ee
as well as the next term in the uniform asymptotic expansion of the
Bessel functions,
\numparts
\bea
\fl s_l(x)\sim
\frac12\sqrt{zt}\,\rme^{\nu\eta}\left(1+u_1(t)/\nu+\dots\right),\qquad
e_l(x)\sim\sqrt{zt}\,\rme^{-\nu\eta}\left(1-u_1(t)/\nu+\dots\right),\\
\fl s'_l(x)\sim\frac12\frac1{\sqrt{zt}}\,\rme^{\nu\eta}\left(1+v_1(t)/\nu+
\dots\right),\qquad
e'_l(x)\sim-\frac1{\sqrt{zt}}\,\rme^{-\nu\eta}\left(1-v_1(t)/\nu+\dots\right),
\eea
\endnumparts
where
$u_1(t)=(3t-5t^3)/24$, $v_1(t)=(3t+7t^3)/24$, as $l\to\infty$.
Including all this, it is straightforward to recover the
well-known result (\ref{divdc}) of Deutsch and Candelas
for strong coupling (Dirichlet BC).
Following the same process for
weak coupling, we find that the 
 leading divergence in order $n$, $1\le n<3$, is ($r\to  a\pm$)
\be
u^{(n)}\sim\left(\frac\lambda{a^2}\right)^{n}
\frac1{1440\pi^2}\frac1{a(a-r)^{3-n}}(n-1)(n+2)\Gamma(3-n).
\ee  
Note that the subleading $\Or(\lambda)$ term again vanishes.
Both of these results  apply 
for the conformal value $\xi=1/6$.

The above results for the conformally coupled scalar
show that the inverse linear divergences which occur in either
order $\lambda$ or $\lambda^2$ cancel between inside and outside, when one
computes the total energy,
while the divergence encountered at $n=3$ is logarithmic:
\be u^{(3)}\sim \frac{\lambda^3}{a^7}\frac1{144\pi^2}\Gamma(0)
\to-\frac{\lambda^3}{144\pi^2 a^7}\ln\frac{r-a}a,
\ee
where the latter form is shown by explicit calculation, rather than
continuing in $n$.
The integral of this, however, is finite, so this does not
signal any difficulty.

\section{Surface and shell  energy}
However, as discussed first by Dowker, Kennedy and Critchley 
\cite{Dowker:1978md,Kennedy:1979ar}, 
and later elaborated by Saharian and Romeo \cite{saharian,romeo},
and put in a broader context by Fulling \cite{fulling}, for situations
when other than Neumann or Dirichlet boundary conditions apply, an
additional term must be supplied in calculating the energy, a term
which resides entirely on the surface.  For the case of the general
stress tensor, that extra term is \cite{Milton:2004vy}
\be
\frak{E}
=-\frac{1-4\xi}{2\rmi}\int \rmd\mathbf{S}\cdot \bnabla G(x,x')\bigg|_{x=x'},
\label{surfterm}
\ee
where the direction of the normal is out of the region in question,
which arises from the $T^{0i}$ component of the stress tensor. 
The total energy in a given region is not, therefore, just the integral
of the local energy density, but has this extra contribution
\cite{Milton:2004vy}
\be
\mathcal{E}=
\int (\rmd\mathbf{r})\langle T^{00}\rangle+\mathfrak{E}=
\frac1{2\rmi}\int(\rmd\mathbf{r})
\int\frac{\rmd\omega}{2\pi}2\omega^2\mathcal{G}(\mathbf{r,r})
\rme^{-\rmi\omega\tau},\label{totalenergy}
\ee
which is independent of $\xi$. ($\tau$ is a point-splitting regulator
\cite{Milton:1978sf}.)
The latter expression has a rather evident interpretation in terms
of summing zero-point energies.  The surface energy cancels for a
nonsingular potential when computing the total energy in all space.

In the limit of $h\to\infty$ for the region in the shell, $a_-<r,r'<a_+$,
the reduced Green's function  becomes (for further details about this 
limit see \cite{surfdiv})
\bea
\fl g_l\to \frac1{2\kappa rr'}\frac{e_l(\kappa a)s_l(\kappa a)}{1+\frac\lambda
{\kappa a^2}e_l(\kappa a)s_l(\kappa a)}
\bigg[\cosh\frac{\sqrt{\lambda h}}a(r-r')
+\cosh\frac{\sqrt{\lambda h}}a(r+r'-a_+-a_-)\bigg].
\eea
In the thin shell limit ($\delta\to0$) this leads to an energy density in the
shell nearly independent of $r$, 
leading to the energy
($\epsilon=\tau_E/a$, $y=|x|$)
\be
E_s=\frac\lambda{4\pi a^2}(1-4\xi)\sum_{l=0}^\infty (2l+1)
\int_{-\infty}^\infty \rmd x
\frac{I_\nu(y)K_\nu(y)}{1+\frac{\lambda}{a} I_\nu(y)K_\nu(y)}
\rme^{\rmi x\epsilon}.\label{shellenergy}
\ee
However, we have to include the surface term 
(\ref{surfterm}) in the shell at $r=a_\pm$,
which exactly cancels this: $\mathcal{E}_s=E_s+\mathfrak{E}_{s}=0$, 
because the total energy
of the shell is given by (\ref{totalenergy}) integrated over the volume
of the shell,
which clearly vanishes as the thickness of the shell $\delta\to0$.
However, we shall see shortly that $E_s$ 
 plays a special role. 

\section{Total energy of $\lambda$ sphere ($\delta=0$)}
Likewise, if one integrates the interior and exterior energy density, 
and includes the surface energy, one gets,
for arbitrary $\xi$,
 the total energy as found by Bordag \etal \cite{Bordag:1998vs},
\be
\fl \mathcal{E}=E_{\rm in}+E_{\rm out}+\mathfrak{E}
=-\frac1{4\pi a}\sum_{l=0}^\infty(2l+1)\int_{-\infty}^\infty\rmd x\,y
\frac{\rmd}{\rmd y}\ln\left[1+\frac{\lambda}{a} I_\nu(y)K_\nu(y) \right]
\rme^{\rmi x\epsilon},\label{totalenergy1}
\ee
exactly that obtained from the integral (\ref{totalenergy})
of the Green's function.


However, there is more to say here.  As noted above, the integral of the
local energy, inside and outside the sphere, is finite perturbatively,
because of cancellations between inside and outside, for the conformally
coupled scalar.
But it is well known that divergences occur in the
total energy at order $\lambda^3$.
These evidently must arise from the
surface term.  So let us consider the latter, which
 is given in  the outside region by
\be
\mathfrak{E}=a^2(1-4\xi)\sum_{l=0}^\infty 2\nu\frac12\int_{-\infty}^\infty 
\frac{\rmd\zeta}{2\pi}\frac{\partial}
{\partial r}g_l(r,r')\bigg|_{r=r'=a}\rme^{\rmi\zeta\tau_E},\qquad 
|\zeta|=\kappa.
\ee
In the strong coupling limit, there is, of course, no surface term.
This is because then
\be
r,r'>a:\quad g_l(r,r')=\frac1{\kappa rr'}\left[s_l(\kappa r_<)e_l(\kappa r_>)
-\frac{s_l(\kappa a)}{e_l(\kappa a)}e_l(\kappa r)e_l(\kappa r')\right],
\ee
which vanishes on the surface, and has a derivative proportional to the
Wronskian.

In general, in the thin-shell limit, the sum of the inside and outside surface 
terms is given by
\be
\mathfrak{E}=\frac\lambda{4\pi a^2}(1-4\xi)\int_{-\infty}^\infty \rmd x 
\sum_{l=0}^\infty (2l+1)
\frac{I_\nu(y)K_\nu(y)}{1+\frac\lambda{a}I_\nu(y)K_\nu(y)}
\rme^{\rmi x\epsilon}.
\ee
Perhaps not remarkably, this is
precisely the same as the integrated local
shell energy $E_s$ (\ref{shellenergy}). 
Thus the surface energies within and outside the shell regions cancel. 
(This is generally true, as follows from the continuity requirements
on the Green's function.)

For weak coupling, we expand this in powers of $\lambda$.
Perhaps the easiest way to isolate the asymptotic behavior is to use
the leading uniform asymptotic expansion,
$I_\nu(x)K_\nu(x)\sim t/2\nu$.
This yields the following expression for the $n$th term in the total
surface energy, ($\epsilon=0$, analytically continued in $n$ from
$\mbox{Re}\,n>3$)
\be
\mathfrak{E}^{(n)}\sim -\frac{(-1)^n}{2\sqrt{\pi}a}(1-4\xi)
\left(\frac{\lambda }{2a}\right)^n
\frac{\Gamma\left(\frac{n-1}2\right)}{\Gamma\left(\frac{n}2\right)}(2^{n-2}-1)
\zeta(n-2).
\ee
 Note that
this expression vanishes for $n=2$; in this approximation
the order $\lambda^2$ term in the energy
arises only from the local energy density. 
 However, for $n=3$ we obtain
for the conformal value, $\xi=1/6$,
\be
\mathfrak{E}^{(3)}\sim \frac{\lambda^3}{24\pi a^4}\zeta(1),
\ee
precisely the divergent term in the energy
first found by the heat kernel calculation of Bordag, Kirsten, and 
Vassilevich \cite{Bordag:1998vs}.
The universality of these results supports the hypothesis of analyticity
in the order $n$.
Alternatively, if we keep $\epsilon\ne0$:
\be \mathfrak{E}^{(2)}\sim -\frac{i\lambda^2}{24\pi\epsilon a^3}
\int_{-\infty}^\infty \frac{dz}z\frac1{z^2+1}=0,\qquad
\mathfrak{E}^{(3)}\sim\frac{\lambda^3}{12\pi a^4}\ln\epsilon.\ee
The former integral vanishes by oddness, while the $\Or(\lambda^3)$ term
is logarithmically divergent as $\epsilon\to0$.  
Thus, by expanding (\ref{totalenergy1}) in powers of $\lambda$,  
$\mathcal{E}^{(2)}=\lambda^2/32\pi a^3$, is
unambiguously finite, while $\mathcal{E}^{(3)}$ is unambiguously 
divergent.


\section{Conclusions}
For the case of a massless scalar field in a spherically symmetric 
step-function shell potential, we have shown that
there is a net effective surface energy in the thin shell
limit, to be added to the 
integrated local energy density for the inside and outside regions, 
which is exactly the integrated local energy density
of the shell. This shell energy, for the conformally coupled
theory, is finite in second order in the
coupling, but diverges in third order.  We show that the latter precisely
corresponds to the known divergence of the total energy in this order.
Thus we have established the suspected correspondence between surface
divergences and divergences in the total energy, which has nothing to
do with divergences in the local energy density as the surface is approached.
This precise correspondence should enable us to absorb such 
global divergences in
a renormalization of the surface energy, and should lead to further advances
of our understanding of quantum vacuum effects.  Further details are
given in \cite{surfdiv}.

\ack
This work was supported by
grants from the U.S. Department of Energy.
We thank Steve Fulling, Noah Graham, Klaus Kirsten,  and Prachi Parashar
 for helpful conversations.  We are grateful to Emilio Elizalde for his
superb organization of the QFEXT05 workshop, and all the participants
of that meeting for making that event such a success.
KAM thanks the Physics Department of Washington University
for its hospitality.

\end{document}